\documentclass[runningheads]{llncs}

\usepackage{comment}
\usepackage{graphicx}
\usepackage{tabularx}
\usepackage{hyperref}
\usepackage{geometry}

\begin{document}

\setlength{\belowcaptionskip}{1pt}
\title{Toward Semantic Publishing in \\ Non-Invasive Brain Stimulation: A Comprehensive Analysis of rTMS Studies\thanks{Supported by German BMBF project SCINEXT (ID 01lS22070), DFG NFDI4DataScience (ID 460234259), and ERC ScienceGraph (ID 819536)}}
\titlerunning{Toward Semantic Publishing in Non-Invasive Brain Stimulation}
%

\author{Swathi Anil\inst{1}\orcidID{0009-0004-5976-1717} \and
Jennifer D'Souza\inst{2}\orcidID{0000-0002-6616-9509}\thanks{Both authors contributed equally to this work.}}

\authorrunning{Anil \& D'Souza, 2023}
%
\institute{University of Freiburg, Germany 
\newline\email{swathianil30@gmail.com}
\and
TIB Leibniz Information Centre for Science and Technology University Library, Germany
\email{jennifer.dsouza@tib.eu}}
\maketitle              
\begin{abstract}

Noninvasive brain stimulation (NIBS) encompasses transcranial stimulation techniques that can influence brain excitability. These techniques have the potential to treat conditions like depression, anxiety, and chronic pain, and to provide insights into brain function. However, a lack of standardized reporting practices limits its reproducibility and full clinical potential. This
paper aims to foster interinterdisciplinarity toward adopting Computer Science Semantic reporting methods for the standardized documentation of Neuroscience NIBS studies making them explicitly Findable, Accessible, Interoperable, and Reusable (FAIR).

In a large-scale systematic review of 600 repetitive transcranial magnetic stimulation (rTMS), a subarea of NIBS, dosages, we describe key properties that allow for structured descriptions and comparisons of the studies. This paper showcases the semantic publishing of NIBS in the ecosphere of knowledge-graph-based next-generation scholarly digital libraries. Specifically, the FAIR Semantic Web resource(s)-based publishing paradigm is implemented for the 600 reviewed rTMS studies in the Open Research Knowledge Graph.


\end{abstract}
\section{Introduction}

Noninvasive brain stimulation (NIBS) is a rapidly evolving field in neuroscience that involves modulating neuronal activity in the brain without surgical intervention, often employing electrical currents or magnetic fields~\cite{peterchev2012fundamentals}. NIBS encompasses various subareas such as transcranial direct current stimulation (tDCS), transcranial alternating current stimulation (tACS), transcranial random noise stimulation (tRNS), repetitive transcranial magnetic stimulation (rTMS), among others. These techniques harness the ability of neural tissue to adapt to external stimulation by making structural, functional and molecular changes, called neural plasticity. Over the past decades, rTMS in particular has gained significant attention for its potential to enhance cognitive abilities, including attention, memory, language, and decision-making. It is also being explored as a treatment option for neuropsychiatric disorders like pharmacoresistant major depressive disorder (MDD), schizophrenia, and obsessive-compulsive disorder, as well as for stroke rehabilitation, pain management, and investigation of brain function~\cite{bikson2001suppression,lopez2010dbs,durand2002break,wachter2011transcranial,karra2010transfection,elwassif2006bio,ardolino2005non,volkow2010effects,martiny2010transcranial,kirson2007alternating}. The outcome of rTMS can be influenced by various stimulation parameters, highlighting the importance of dosage considerations. Stimulation parameters such as intensity, duration, frequency, and pattern play a significant role in shaping the effects of rTMS on brain activity and clinical outcomes.

An ongoing concern in the field of repetitive transcranial magnetic stimulation (rTMS) is the lack of standardized reporting for dosing parameters, which has implications for comprehensive understanding and clinical application of these techniques~\cite{wilson2016repetitive}. Currently, the reporting of rTMS dosage relies on textual descriptions, leading to varying levels of detail and specificity. This variability hampers reproducibility and the establishment of consistent protocols and effective stimulation parameters.
The absence of standardized reporting practices for rTMS dosage poses challenges in comparing and replicating studies, limiting the accumulation of robust evidence and the generalizability of findings across different populations. This also poses a disadvantage in the ability to conduct meta-analyses, hindering the translation of rTMS into clinical practice in an effective manner. Therefore, efforts to establish standardized reporting guidelines for rTMS studies are warranted, aiming to enhance transparency, reproducibility, and collaboration within the field.

In this context, we recognize the opportunity for interdiciplinarity. In Computer Science, the semantic web publishing method utilizes semantic technologies and ontologies to structure and standardize data, enabling more efficient and meaningful reporting, querying, and analysis of information within the World Wide Web~\cite{berners2001semantic}. We thus posit the adoption of semantic web methods from Computer Science as a solution to the problem of non-standardized reporting of Neuroscience NIBS study doses. This is in keeping with the recent growing general impetus driving a change in the status quo of discourse-based scholarly communication toward Findable, Accessible, Interoperable, and Resuable (FAIR)~\cite{wilkinson2016fair} publishing as structured scholarly knowledge graphs by adopting the technologies of the semantic web~\cite{scigraph,openaire,researchgraph,baas2020scopus,birkle2020web,wang2020microsoft,hendricks2020crossref,auer2020improving}. We claim that \textit{the adoption of FAIR reporting paradigms facilitated by semantic web technologies for NIBS dosing will mark a significant stride toward overcoming the pressing challenge of reproducibility in this field.} In turn, it will facilitate tapping into the full clinical potential of NIBS. 

\begin{figure}[!htb]
\includegraphics[width=\textwidth]{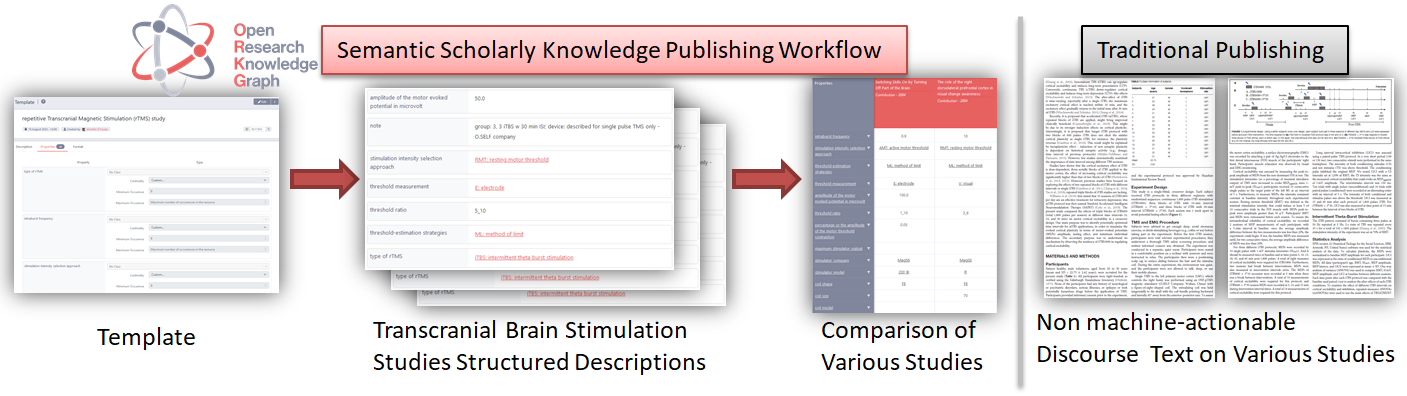}
\caption{Machine-actionable structured scholarly knowledge capture via semantic publishing (in red) versus traditional discourse-based non-machine-actionable publishing (in gray).}
\label{publishing-workflow}
\end{figure}

Toward the semantic web publishing of NIBS dosages a few technical considerations need to be in place. These include: defining the salient entities, attributes, and relationships for the NIBS domain as standardized resources accessible on the Web via uniform resource identifiers (URIs); optionally, designing or reusing a suitable ontology that captures the concepts, relationships, and constraints of interest; and publishing NIBS doses using structured triples conforming to the RDF syntax recommended by the World Wide Web Consortium (W3C).\footnote{\url{https://www.w3.org/TR/rdf11-concepts/}} It is crucial to utilize standard ontologized vocabulary terms, incorporate appropriate metadata, and ensure data publication on the web with unique URIs for resource identification and interlinking. Fortunately, the latter required FAIR semantic web publishing functionalities are readily available within next-generation scholarly knowledge publishing digital libraries such as the Open Research Knowledge Graph (ORKG)~\cite{auer2020improving,stocker2023fair}. The ORKG platform supports the semantic publishing~\cite{shotton2009semantic} of scholarly knowledge as structured, machine-actionable data. Its specific functionalities include offering a robust and reliable web namespace within which to specify the vocabulary of pertinent scholarly knowledge via dereferenceable uniform resource identifiers (URIs); intuitive frontend interfaces~\cite{oelen2022leveraging} to define scholarly resources,\footnote{\url{https://orkg.org/resources}} properties,\footnote{\url{https://orkg.org/properties}} classes,\footnote{\url{https://orkg.org/classes}} and templates\footnote{\url{https://orkg.org/templates}} which via the platform's backend technologies automatically represent the growing scholarly knowledge graph in RDF syntax;\footnote{\url{https://orkg.org/api/rdf/dump}} given the machine-actionability that semantic representations of knowledge generally entail, the ORKG supports the creation of novel comparison views over comparably structured scholarly knowledge in a similar way to comparisons of products on e-commerce websites~\cite{oelen2020}; and finally, SPARQL queryable interfaces\footnote{\url{https://orkg.org/triplestore}, \url{https://orkg.org/sparql/}} to further obtain customized snapshots or aggregated views of the ORKG are also supported. \autoref{publishing-workflow} juxtaposes the  semantic publishing of scholarly knowledge in the ORKG in terms of one of its workflows compared with traditional discourse-based scholarly publishing paradigm.

\begin{figure}
\centering
\includegraphics[width=\textwidth]{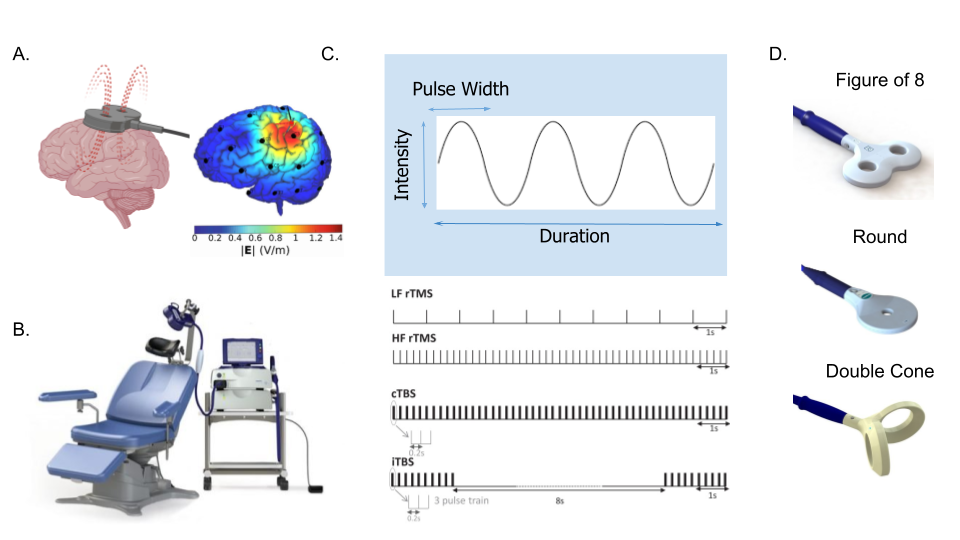}
\caption{Repetitive Transcranial Magnetic Stimulation (rTMS): \textbf{A.} Schematic representation of an rTMS coil positioned over a brain. The illustration shows the precise placement of the coil, adhering to the specific anatomical target, alongside corresponding computational estimation of the electric field (E-field) generated by the rTMS coil in the brain. The color gradient indicates the intensity of the electric field, highlighting the spatial distribution of the stimulation. \textbf{B.} Conventional rTMS device and experimental setup. \textbf{C.} Parameters that can influence rTMS outcome include pulse width, frequency, intensity and net protocol duration, among others. \textbf{D.} rTMS coil shapes can influence the electrical field distribution of stimulation over the brain, impacting the stimulation outcome. Courtesy: \href{https://www.biorender.com/}{Biorender.com}, \href{https://simnibs.github.io/simnibs/build/html/index.html}{SIMNIBS}, Magstim.}
\label{TMS}
\end{figure}

Returning then to the interdisciplinary vision laid out in this paper toward adopting the semantic web publishing methodology for NIBS dosages, the only remaining action items would then be to define the standardized vocabulary to represent the domain. With this goal in mind, in this paper, we showcase in practice the semantic publishing of repetitive transcranial magnetic stimulation (rTMS) -- a subarea of NIBS -- as a use-case toward the semantic publishing of NIBS studies, generally. rTMS is a non-invasive neuromodulation technique that applies magnetic pulses to targeted brain regions to modulate neural activity~\cite{wassermann2001therapeutic}. Extensive research in both humans and animals has demonstrated the therapeutic potential of rTMS in neurological and psychiatric disorders. Notably, rTMS has shown efficacy in conditions such as major depressive disorder (MDD) and obsessive-compulsive disorder (OCD). \autoref{TMS} offers an illustrative overview of the rTMS methdology in practice. The U.S. Food and Drug Administration (FDA) has granted approval for the use of rTMS as a therapeutic tool, specifically for the treatment of MDD and OCD. However, due to the non-standardization of experimental protocols and data reporting of rTMS, the evidence to inform clinical application is highly inconsistent~\cite{thut2010review,hamada2013role} and
substantially based on trial and error. As a first step, is the discovery of standardized vocabulary to describe rTMS. An early work in this direction is the comprehensive review of rTMS studies between 1999 and 2020 by Turi et al.~\cite{turi2021selecting} which introduced a set of salient properties based on which over 600 rTMS studies were comparable. The standardized applicability of the properties over the large number of studies indicates their potential to be realized as a schema based on the philosophy of data-driven bottom-up ontology construction~\cite{van1998bottom}. Thus, in this paper we investigate the following main research questions (\textbf{RQ}s). \textbf{RQ1}: What are the properties for the structured representation of rTMS studies? This is addressed in \autoref{semantification}. \textbf{RQ2}: How can this semantic reporting of rTMS studies specifically, and NIBS studies generally, be practically realized as a FAIR model? -- addressed in \autoref{FAIR}.



The primary goal of this work is to engage a broad and varied audience, especially in the field of neuroscience. We aim to foster interdisciplinary discussions to enhance understanding of the fast-evolving topic of semantic scholarly knowledge publishing, both within this paper's context and beyond. The proposed semantics-based structured recording method is a concrete first step that directly addresses the two main long-discussed problems of \textit{transparency} and \textit{reproducibility} of NIBS studies to facilitate clinical application~\cite{wilson2016repetitive}. Our concrete implementation of the semantic web-based technology reporting of rTMS studies serves as a demonstration use-case and a call-to-action to adopt and foster collaborations toward ORKG infrastructural extensions as a platform to support publishing of NIBS studies.


\section{Background on rTMS Dosage Data Management}

The management of data in rTMS studies remains a challenging task due to the varied practices in data collection, storage, and protection protocols \cite{wilson2016repetitive}. Data is commonly collected directly from physiological signals and stored in localized databases or documentation tools. While these practices fulfil basic data storage needs, they raise significant issues concerning data interoperability, reusability, and long-term accessibility, primarily due to inconsistencies in data structures and metadata provision \cite{wilson2016repetitive}. Moreover, data protection presents a significant challenge in this context. The sensitivity of personal data demands stringent measures to ensure security, which can further complicate data accessibility. Ensuring secure access requires a fine balance between protection and availability, often calling for robust authentication protocols and persistent identifiers \cite{wilkinson2016fair}. Adding to the complexities is the variability in the selection of stimulation parameters, in particular the stimulation intensity, across rTMS studies, as highlighted in Turi et al.'s systematic review \cite{turi2021selecting}. The lack of a standardized protocol for this crucial parameter leads to substantial inconsistencies across studies, impacting the reliability and comparability of research findings.

To refine this process, the integration of FAIR (Findable, Accessible, Interoperable, and Reusable) principles into rTMS dosage data management presents a potential solution \cite{wilkinson2016fair}. With the application of these principles, data could be catalogued accurately, improving findability. Accessibility could be ensured with robust authentication protocols, allowing for a more secure and open sharing of data. Interoperability, although often not prioritized in traditional rTMS studies, could be significantly enhanced through common data models and agreed-upon formats \cite{bierer2017}. The proposed adaptations in rTMS data handling, encompassing FAIR principles and standardized intensity selection, can substantially elevate the quality of research outputs, making them more robust, reliable, and universally acceptable.

\begin{table}
\centering
\renewcommand{\arraystretch}{1.3}
\caption{rTMS key properties and associated resources}
\label{tab:Table1}
\footnotesize
\begin{tabularx}{\textwidth}{|X|X|}
\hline
\textbf{Property} & \textbf{Resources} \\
\hline
Type of rTMS & Conventional rTMS (rTMS), Intermittent theta burst stimulation (iTBS), Continuous theta burst stimulation (cTBS), Quadripulse stimulation (QPS) \\
\hline
Intrabust Frequency & - \\
\hline
Stimulation Intensity Selection Approach & Active motor threshold (AMT), Resting motor threshold (RMT), Unspecified motor threshold (MT), Functional lesion (FL), Phosphene threshold (PT), Fixed intensity (FXD), Electric field (EF) \\
\hline
Threshold-estimation strategies & Method of limit (ML), 5 step procedure (5STEP), Threshold hunting (TH), Maximum likelihood based threshold hunting (MLTH), Parameter estimation by sequential testing (PEST), TMS Motor Threshold Assessment Tool (MTAT) \\
\hline
Threshold Measurement & Electrode (E), Visual (V) \\
\hline
Amplitude of the Motor Evoked Potential (mV) & - \\
\hline
Threshold Ratio & - \\
\hline
Percentage or the Amplitude of the Motor Threshold Contraction & - \\
\hline
Percent of Stimulation Intensity & Min value, Max value \\
\hline
Maximum Stimulator Output & - \\
\hline
Stimulator Company & Cad, MedDan, MagSti, NeoNet, NeuNet, MagVen, NexSti, MagMor, Yir, BraSwa, DeyDia, YunTec, NeuSof \\
\hline
Stimulator Model & HS, MP, MES10, R, SR, SR2, NP, 16E05, 200, 200 2, MLR25, 200 BI, QP500, HF, MP30, MPX100, 2100CRS, MP100, R2, MPR30, NBS, PM100, CCYI, CCYIA, DMXT, NS, Sys4.3, R2P1, N-MS/D, MPC, MS/D \\
\hline
Coil Shape & F8, R, F8-D, D \\
\hline
Coil Size & - \\
\hline
Coil Model & MC125, MC-125, MC-B70, MCF-B70, MCF-B-65, MCF-B65, WC, AC, DC, PN9925, 992500, C-B60, FC, FC-B70, HP, Cool B65, cool-B65, Cool-DB80, Cool B56, H-ADD, H, H1, AF, DB-80, B65, MMC-140, 70BF-Cool \\
\hline
\end{tabularx}
\end{table}

\section{Supporting Semantic Technology and Semantification of rTMS Studied Doses}
\label{semantification}

As a first step, for the semantic publishing of rTMS dosage, we need to define the rTMS-domain-specific resources and properties. Using the ORKG semantic scholarly knowledge publishing platform, we decide to define the rTMS resources and predicates as resources on the web with deferenceable URIs in the ORKG namespace. The use of the ORKG namespace offers: \textit{1. Standardization and Interoperability.} It provides a framework for representing scholarly knowledge, ensuring seamless integration and data sharing across platforms. \textit{2. Community Collaboration.} Researchers and practitioners can actively contribute to the growth and improvement of the ORKG ecosystem, refining ontology and shaping scholarly resources. \textit{3. Enhanced Discoverability.} Adoption of the ORKG namespace improves the visibility and indexing of scholarly resources, increasing their chances of being accessed and cited. \textit{4. Semantic Enrichment.} The comprehensive ontology of the ORKG namespace enriches scholarly resources with detailed metadata, facilitating advanced queries and analysis. \textit{5. Community Trust and Credibility.} Utilizing the ORKG namespace adds credibility to scholarly resources through community validation, fostering trust within the scholarly community. \textit{6. Long-term Sustainability.} The ORKG namespace, supported by an active community, ensures the maintenance and longevity of semantic web resources, giving researchers confidence in their continuity. Overall, the use of the ORKG namespace provides researchers and organizations with a standardized, collaborative, and credible framework for creating scholarly resources and predicates in the semantic web. 

rTMS doses encompass multiple adjustable parameters that shape the characteristics of the stimulation~\cite{lefaucheur2014evidence,lefaucheur2020evidence,rossi2009safety,wassermann1998risk,miranda2006modeling,thielscher2015field}. The longitudinal review of rTMS studies between 1999 and 2020 by Turi et al.~\cite{turi2021selecting} characterized the rTMS dose in terms of 15 salient properties. Based on the philosophy of bottom-up discovery of standardized domain vocabulary~\cite{van1998bottom}, these 15 properties and their resources are defined as URIs in the ORKG namespace and then applied to document the individual rTMS doses reviewed. The 15-properties strong rTMS recording resources, detailed in Table \ref{tab:Table1}, are: \textbf{1.} \href{https://orkg.org/property/P78000}{\textit{type of rTMS}} with four resource candidates, viz. conventional rTMS (\href{https://orkg.org/resource/R254150}{rTMS}), intermittent theta burst stimulation (\href{https://orkg.org/resource/R254175}{iTBS}), continuous theta burst stimulation (\href{https://orkg.org/resource/R254154}{cTBS}), and quadripulse stimulation (\href{https://orkg.org/resource/R254176}{QPS}). \textbf{2.} \href{https://orkg.org/property/P78001}{\textit{intrabust frequency}}. \textbf{3.} \href{https://orkg.org/property/P78002}{\textit{stimulation intensity selection approach}} with seven resource candidates, viz. active motor threshold (\href{https://orkg.org/resource/R254177}{AMT}), resting motor threshold (\href{https://orkg.org/resource/R254178}{RMT}), unspecified motor threshold (\href{https://orkg.org/resource/R254179}{MT}), functional lesion (\href{https://orkg.org/resource/R254180}{FL}), phosphene threshold (\href{https://orkg.org/resource/R254181}{PT}), fixed intensity (\href{https://orkg.org/resource/R254182}{FXD}), and electric field (\href{https://orkg.org/resource/R254183}{EF}). \textbf{4.} \href{https://orkg.org/property/P78003}{\textit{threshold-estimation strategies}} with six resource candidates, viz. method of limit (\href{https://orkg.org/resource/R254301}{ML}), 5 step procedure (\href{https://orkg.org/resource/R254302}{5STEP}), threshold hunting (\href{https://orkg.org/resource/R254303}{TH}), maximum likelihood based threshold hunting (\href{https://orkg.org/resource/R254304}{MLTH}), parameter estimation by sequential testing (\href{https://orkg.org/resource/R254305}{PEST}), and TMS Motor Threshold Assessment Tool (\href{https://orkg.org/resource/R254306}{MTAT}). \textbf{5.} \href{https://orkg.org/property/P78005}{\textit{threshold measurement}} with two resource candidates, viz. electrode (\href{https://orkg.org/resource/R254307}{E}) and visual (\href{https://orkg.org/resource/R254308}{V}). \textbf{6.} \href{https://orkg.org/property/P78006}{\textit{amplitude of the motor evoked potential in microvolts}}. \textbf{7.} \href{https://orkg.org/property/P78007}{\textit{threshold ratio}}. \textbf{8.} \href{https://orkg.org/property/P78008}{\textit{percentage or the amplitude of the motor threshold contraction}}. \textbf{9.} \href{https://orkg.org/property/P114005}{\textit{percent of stimulation intensity}} with sub-properties 9.a. \href{https://orkg.org/property/P114006}{\textit{percent of stimulation intensity (min value)}} and 9.b. \href{https://orkg.org/property/P114007}{\textit{percent of stimulation intensity (max value)}}. \textbf{10.} \href{https://orkg.org/property/P78009}{\textit{maximum stimulator output}}. \textbf{11.} \href{https://orkg.org/property/P78010}{\textit{stimulator company}} with 13 and counting resource candidates: \href{https://orkg.org/resource/R254309}{Cad}, \href{https://orkg.org/resource/R254310}{MedDan}, \href{https://orkg.org/resource/R254311}{MagSti}, \href{https://orkg.org/resource/R254312}{NeoNet}, \href{https://orkg.org/resource/R254313}{NeuNet}, \href{https://orkg.org/resource/R254314}{MagVen}, \href{https://orkg.org/resource/R254315}{NexSti}, \href{https://orkg.org/resource/R254316}{MagMor}, \href{https://orkg.org/resource/R254317}{Yir}, \href{https://orkg.org/resource/R254318}{BraSwa}, \href{https://orkg.org/resource/R254319}{DeyDia}, \href{https://orkg.org/resource/R254320}{YunTec}, and \href{https://orkg.org/resource/R254321}{NeuSof}. \textbf{12.} \href{https://orkg.org/property/P78011}{\textit{stimulator model}} with 31 resources: \href{https://orkg.org/resource/R254322}{HS}, \href{https://orkg.org/resource/R254323}{MP}, \href{https://orkg.org/resource/R254324}{MES10}, \href{https://orkg.org/resource/R254325}{R}, \href{https://orkg.org/resource/R254326}{SR}, \href{https://orkg.org/resource/R254327}{SR2}, \href{https://orkg.org/resource/R254328}{NP}, \href{https://orkg.org/resource/R254329}{16E05}, \href{https://orkg.org/resource/R254330}{200}, \href{https://orkg.org/resource/R254331}{200\_2}, \href{https://orkg.org/resource/R254332}{MLR25}, \href{https://orkg.org/resource/R254333}{200\_BI}, \href{https://orkg.org/search/QP500}{QP500}, \href{https://orkg.org/resource/R254335}{HF}, \href{https://orkg.org/resource/R254336}{MP30}, \href{https://orkg.org/resource/R254337}{MPX100}, \href{https://orkg.org/resource/R254338}{2100CRS}, \href{https://orkg.org/resource/R254339}{MP100}, \href{https://orkg.org/resource/R254340}{R2}, \href{https://orkg.org/resource/R254341}{MPR30}, \href{https://orkg.org/resource/R254342}{NBS}, \href{https://orkg.org/resource/R254343}{PM100}, \href{https://orkg.org/resource/R254344}{CCYI}, \href{https://orkg.org/resource/R254345}{CCYIA}, \href{https://orkg.org/resource/R254346}{DMXT}, \href{https://orkg.org/resource/R254347}{NS}, \href{https://orkg.org/resource/R254348}{Sys4.3}, \href{https://orkg.org/resource/R254349}{R2P1}, \href{https://orkg.org/resource/R254350}{N-MS/D}, \href{https://orkg.org/resource/R254351}{MPC}, and \href{https://orkg.org/resource/R254352}{MS/D}. \textbf{13.} \href{https://orkg.org/property/P78012}{\textit{coil shape}} with 4 resources: \href{https://orkg.org/resource/R254353}{F8}, \href{https://orkg.org/resource/R254354}{R}, \href{https://orkg.org/resource/R254355}{F8-D}, and \href{https://orkg.org/resource/R254356}{D}. \textbf{14.} \href{https://orkg.org/property/P78013}{\textit{coil size}}. \textbf{15.} \href{https://orkg.org/property/P78014}{\textit{coil model}} with 27 resources: \href{https://orkg.org/resource/R254357}{MC125}, \href{https://orkg.org/resource/R254358}{MC-125}, \href{https://orkg.org/resource/R254359}{MC-B70}, \href{https://orkg.org/resource/R254360}{MCF-B70}, \href{https://orkg.org/resource/R254361}{MCF-B-65}, \href{https://orkg.org/resource/R254362}{MCF-B65}, \href{https://orkg.org/resource/R254363}{WC}, \href{https://orkg.org/resource/R254364}{AC}, \href{https://orkg.org/resource/R254365}{DC}, \href{https://orkg.org/resource/R254366}{PN9925}, \href{https://orkg.org/resource/R254367}{992500}, \href{https://orkg.org/resource/R254368}{C-B60}, \href{https://orkg.org/resource/R254369}{FC}, \href{https://orkg.org/resource/R254370}{FC-B70}, \href{https://orkg.org/resource/R254371}{HP}, \href{https://orkg.org/resource/R254372}{Cool B65}, \href{https://orkg.org/resource/R254373}{cool-B65}, \href{https://orkg.org/resource/R254374}{Cool-DB80}, \href{https://orkg.org/resource/R254375}{Cool B56}, \href{https://orkg.org/resource/R254376}{H-ADD}, \href{https://orkg.org/resource/R254377}{H}, \href{https://orkg.org/resource/R254378}{H1}, \href{https://orkg.org/resource/R254379}{AF}, \href{https://orkg.org/resource/R254380}{DB-80}, \href{https://orkg.org/resource/R254381}{B65}, \href{https://orkg.org/resource/R254382}{MMC-140}, and \href{https://orkg.org/resource/R254383}{70BF-Cool}. These properties are implemented as a template \url{https://orkg.org/template/R211955} which can be extended and reused for the explicit and standardized recording of rTMS doses. As a demonstration use-case, the 600 reviewed studies~\cite{turi2021selecting} were semantically published resulting in an rTMS-KG as a subgraph of the ORKG. Comparisons over the structured rTMS data were computed and published in six parts, i.e. \href{https://orkg.org/comparison/R589436/}{\underline{1}}, \href{https://orkg.org/comparison/R589474/}{\underline{2}}, \href{https://orkg.org/comparison/R589496/}{\underline{3}}, \href{https://orkg.org/comparison/R589505/}{\underline{4}}, \href{https://orkg.org/comparison/R589510/}{\underline{5}}, \& \href{https://orkg.org/comparison/R589518/}{\underline{6}}, with $\sim$100 studies compared in each part.

\subsection{rTMS-KG - Findable, Accessible, Interoperable, and Reusable}
\label{FAIR}

rTMS-KG is an open-source machine-actionable knowledge graph of 600 rTMS doses that satisfies the FAIR publishing guidelines~\cite{wilkinson2016fair}. Each published comparison of studies in the rTMS-KG can be persistently identified with a Digital Object Identifier (DOI) (`Findable'). This resolves to a landing page, providing access to metadata, respective rTMS study data, and versioning, all of which is indexed and searchable (‘Findable’, ‘Accessible’, and ‘Reusable’). rTMS-KG can be downloaded over the Web from the ORKG in RDF (‘Interoperable’). The rTMS vocabulary can be mapped to existing relevant ontologies via same-as links (‘Interoperable’). Finally, the ORKG provides public search interfaces to build custom views of rTMS-KG (‘Accessible’).

\section{Conclusion}
Standardization of experimental protocols and data reporting is pivotal for the advancement of the rTMS field in particular, and NIBS more broadly. Adopting consensus reporting guidelines and protocols would foster uniformity and comparability across studies, thereby facilitating a more precise evaluation of NIBS dose efficacy. Transparent reporting practices, coupled with data sharing, can bolster reproducibility, reduce publication bias, and encourage collaboration. This enhancement can be directly achieved through the semantic web publishing of scholarly knowledge, as exemplified by the ORKG. By defining a standardized vocabulary of properties and resources, and documenting upcoming studies within this predefined semantic framework, the issues of \textit{transparency} and \textit{reproducibility} in NIBS studies are addressed. As a result, the NIBS field can continue its trajectory of progress, delivering effective, evidence-based neuromodulation interventions for neurological and psychiatric disorders~\cite{wilson2016repetitive}.

This study, for the first time, showcases the practical application of the semantic web resource-based reporting methodology for NIBS studies at large, and rTMS studies in particular. It offers transformative value by: (1) promoting standardization in the recording of NIBS enhances transparency through the explicit logging of key study properties, which in turn bolsters reproducibility, and (2) promoting the adoption of such machine-actionable documentation for NIBS studies, as evidenced by the demonstrated rTMS subarea use-case.

\section*{Acknowledgements}

We thank the anonymous reviewers for their detailed and insightful comments on an earlier draft of the paper. This work was jointly supported by the German BMBF project SCINEXT (ID 01lS22070), DFG NFDI4DataScience (ID 460234259), and ERC ScienceGraph (ID 819536). 

%
%
%
\bibliographystyle{splncs04}
\bibliography{references}

\end{document}